\UseRawInputEncoding
\documentclass{sf2a-conf2021}
\usepackage{graphicx}
\usepackage{hyperref}
\usepackage[]{natbib}  
\usepackage{epstopdf}
\usepackage[T1]{fontenc}
\usepackage[french]{babel}
\usepackage{caption}
\usepackage{hyphenat}
\def\BibTeX{{\rm B\kern-.05em{\sc i\kern-.025em b}\kern-.08em
    T\kern-.1667em\lower.7ex\hbox{E}\kern-.125emX}}
\bibpunct{(}{)}{;}{a}{}{,}  

\begin{document}

\TitreGlobal{SF2A 2021}


\title{Well-being in French astrophysics}

\runningtitle{Well-being in French astrophysics}

\author{N. A. Webb}\address{IRAP, Universit\'e de Toulouse, CNRS, CNES, Toulouse, France}

\author{C. Bot}\address{CDS, Universit\'e de Strasbourg, CNRS, Observatoire astronomique de Strasbourg, UMR 7550, 67000 Strasbourg, France}
\author{S. Charpinet$^1$}
\author{T. Contini$^1$}
\author{L. Jouve$^1$}
\author{F. Koliopanos$^1$}
\author{A. Lamberts}\address{Universit\'e C\^ote d'Azur, Observatoire de la C\^ote d'Azur, CNRS, Laboratoire Lagrange, Laboratoire Art\'mis, France}
\author{H. Meheut$^3$}
\author{S. Mei}\address{Universit\'e de Paris, CNRS, Astroparticule et Cosmologie, F-75013 Paris, France}
\author{I. Ristorcelli$^1$}
\author{G. Soucail$^1$}




\setcounter{page}{237}


\maketitle


\begin{abstract}
It has become clear that early career astrophysics researchers (doctoral researchers, post-docs, etc) have a very diverse appreciation of their
career, with some declaring it the best job that you can have and
others suffering from overwork, harrassment and stress from the precarity of their job, and associated difficulties. In order to establish
how astrophysics researchers, primarily in France, experience their
career, we sent out a survey to understand the impact that their job
has on their well-being. 276 people responded to the survey. Whilst
around half of the respondents expressed pleasure derived from their
career, it is clear that many (early career) researchers are suffering
due to overwork, with more than a quarter saying that they work in excess of
50 hours per week and 2\% in excess  of 90 h per week. Almost 30\% professed to having suffered harrassment or discrimination in the course of their work. Further, whilst only 20\% had suffered mental health issues before starting their career in astrophysics, $\sim$45\% said that they suffered with mental health
problems since starting in astrophysics.  Here we provide results from the survey as well as possible avenues to explore and a list of recommendations to improve (early) careers in astrophysics.

\end{abstract}



\section{Introduction}

Astrophysics is an exciting subject that attracts many young people, as there are many rewards to be had in research in this field, such as enjoying intellectual challenges, interacting with knowledgeable and fascinating colleagues, the possibility to be creative and discover new things and travelling, to name but a few. However, in recent years a general feeling of discouragement has been observed in a significant fraction of doctoral researchers and post-docs \citep[e.g.][]{wool19,auer18}. In some, the feelings are tending towards distress, which can lead to terrible consequences.

To understand the situation in France, we put in place an anonymous survey using the platform {\em Framaforms}. The survey was open from 29th March - 3rd May 2021 and the questions were provided in English and French. It was announced via the French national astronomy and astrophysics society (SF2A) newsletter, in French astrophysics laboratories, as well as a couple of Swiss/Canadian institutes and some French nuclear and particle physics institutes with groups working on astrophysics. The questionnaire covered five main areas, namely general questions about work, such as favourite aspects of astrophysics research, the number of hours worked, perceived external constraints and future plans for remaining in the field. We also asked about relationships with colleagues, notably any issues concerning harassment or discrimination. We questioned about mental health issues, both before and since working in astrophysics. We requested suggestions for improvements that could be made as well as positive feedback and finally we inquired about some demographics, such as the respondents age, position, country where they are working, nationality and gender. None of the questions were obligatory. The results and a discussion are presented in the following sections, taking into account the small number statistics and eventual bias. Our aim was to identify reasons for any suffering, identify improvements that can be made, propose solutions to laboratories, doctoral schools and gouverning bodies and ultimately improve the PhD, post-doc and career experience.
  
\section{Results from the survey}
\label{sec:results}

276 people responded to  the survey, but as none of the questions were obligatory, not everyone responded to each question. 57.8\% of respondents identified as male, 40.6\% as female and 1.6\% identified as other. The survey was open to anyone from planetary science/astrophysics or connecting areas, but targeted doctoral researchers and post-docs. As a result, the majority of the respondents were doctoral researchers (108 respondents) and post-docs (99 respondents). 9 respondents were undergraduates and 62 in a permanent position (27 with (teaching) duties and 35 with no formal duties). The large majority of respondents (232) worked in France, with only 46 working outside of the France. Two thirds (190 people) were French and one third (86 people) were of a different nationality.

When asked about what was the best aspect of the job, almost all respondents replied that they enjoyed the intellectual challenge (243 people), discovering new things (209), the independence they had in their work (192) and interacting with colleagues (188). About half of the respondents enjoyed the creativity of their work (147 respondents), travelling (119) and sharing their discoveries (114 people). Less people cited the University/academic environment as their preferred aspect of their job (97 people), 54 appreciated the fact that their acquired skills will be useful for a future position, 39 cited the social life and just 6 evoked their salary as one of their favourite things. Other things that were occasionally cited were the prestige of the position, the international environment and supervising students.

\begin{figure}[ht!]
 \centering
 \includegraphics[width=0.65\textwidth,clip]{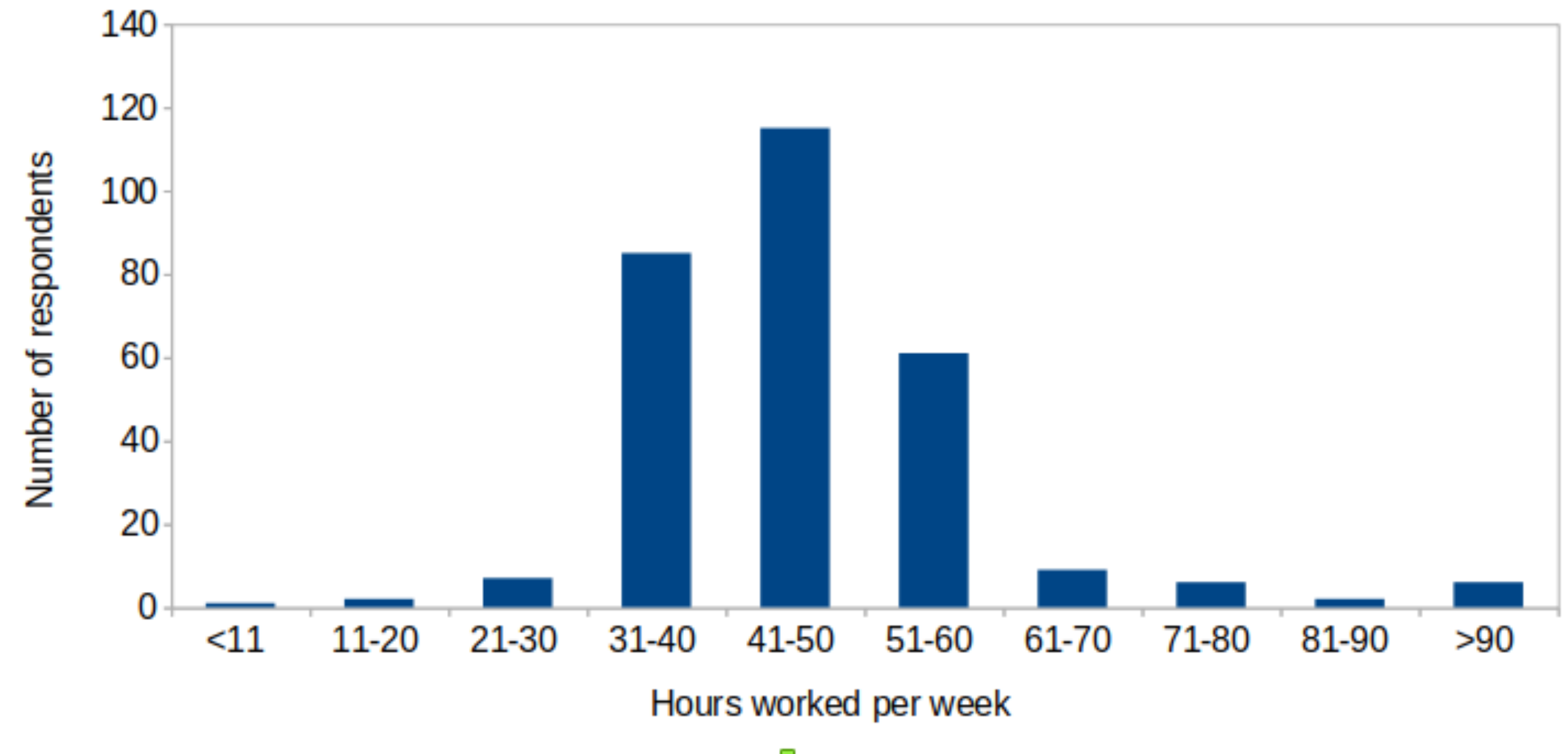}      
  \caption{The number of hours worked per week by the respondents.}
  \label{fig:hours_worked}
\end{figure}

Figure~\ref{fig:hours_worked} shows the number of hours worked per week estimated by the respondents. In France, the working week is 35 hours and doctoral researchers sign a contract stipulating this volume. 44\% of respondents felt external pressure to work outside legal hours, whilst 49\% did not feel that pressure (7\% did not know whether they felt external pressure). 55\% felt that they should work outside of legal hours (37\% that they shouldn't and 8\% did not know). The reasons cited as to why long hours were worked were to be competitive/obtain a permanent position, because they enjoyed working more hours, because they can't achieve the work required without working longer hours, they are unable to {\em switch off}, they felt that more work is expected from post-doc/PhDs than permanent staff and some felt that as others work outside legal hours, so should they.

With regards to job satisfaction, more than half agreed with the statement {\em Research is the best job that you can have} and almost three quarters felt that {\em their work gave them a sense of purpose} and just over a half agreed that {\em they looked forward to going to work every day}. 61\% felt {\em well integrated in their institute}.

With regards to work-life balance, 62\% said that they got sufficient sleep most nights. 35\% felt that their life was balanced with respect to work, outside activities and sleep. 55\% were satisfied with their living conditions and 48\% were satisfied with their financial situation.  54\% felt that the environment in which they work inspires them, 61\% agreed that senior colleagues are there for them when needed and 48\% agreed that they handle setbacks and disappointments in their work well. Just 36\% agreed that they were satisfied with their career progression. 64\% planned to stay in academia, with 11\% planning to leave and 25\% unsure. However, when asked if they would stay in academia if they were guaranteed a permanent position, 81\% said they would stay, 5\% said that they would leave and 14\% were unsure.  Difficulties cited were precarity (most highly cited), the geographical instability due short term contracts, the pressure felt to publish both frequently and high impact papers, feeling undervalued, the general disregard for well-being in the domain, supervisors with (very) poor management skills, difficulties in competing with child-free colleagues, a perception that people in positions of power may misuse their privilege, a lack of information provided for non-academic areas of research (how to access resources, etc) and the large amount of bureaucracy, especially when entering a new institute.

With regards to people's living situations, 13 people said that they had taken on a second job to supplement their salary and 30 people were struggling with debt. 23 post-docs had children or family members to care for, while 6 doctoral researchers had family members to care for. 20 doctoral researchers and 35 postdocs were living apart from their partners and/or children. 10 permanent staff were also living apart from their partners and/or children. 30\% of the women were living apart from their partners and/or children but only 18\% of men.

\begin{figure}[ht!]
 \centering
 \includegraphics[width=0.65\textwidth,clip]{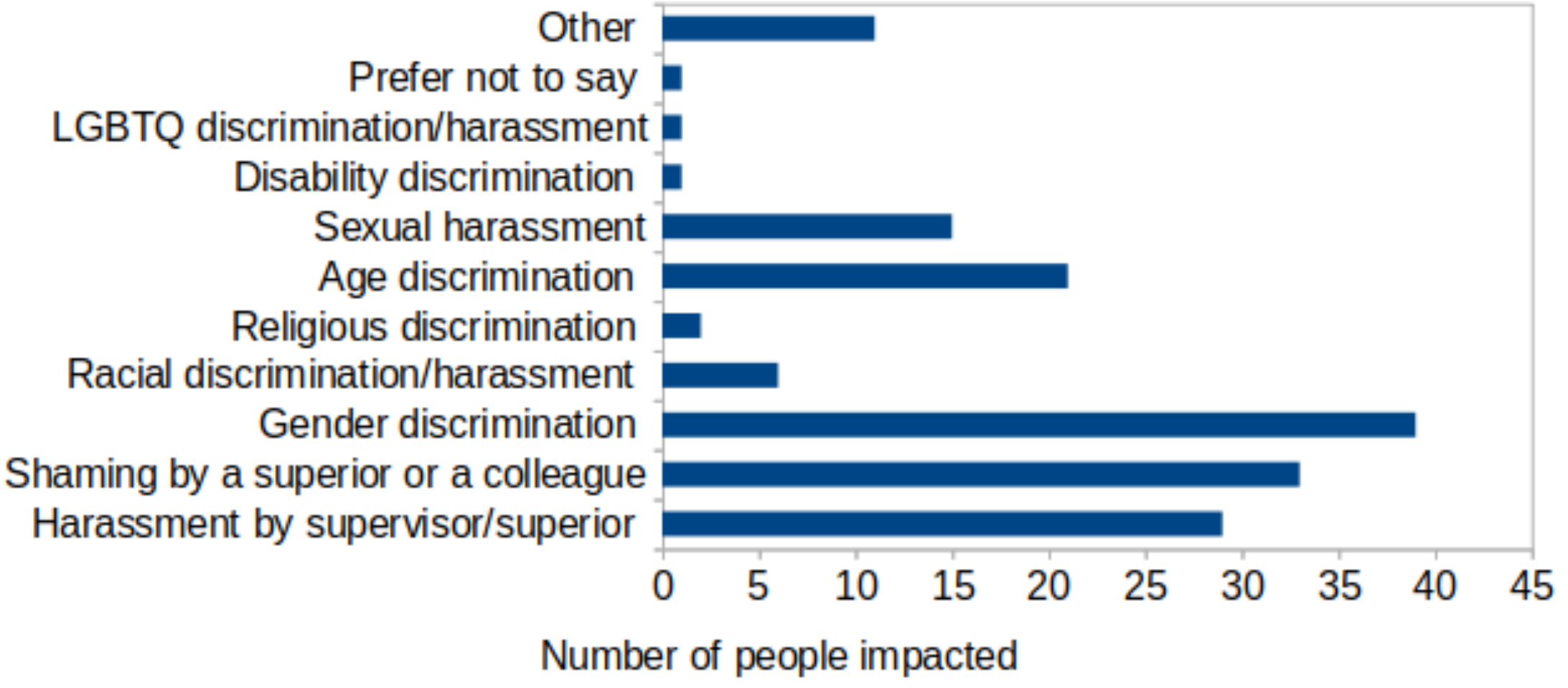}      
  \caption{The nature of the different types of harassment/discrimination encountered by respondents and the number of people experiencing the different types of incidents. Respondents who chose {\em other} cited discrimination based on nationality, hierarchy, due to having kids (or not), language discrimination and discrimination due to research subject.
}
  \label{fig:harassment}
 \end{figure}

29\% of the people responding to the survey had experienced harassment or discrimination since working in the domain. A further 4\% were unsure whether they had experienced harassment or discrimination. Only 26\% of those that had experienced harassment or discrimination reported the incident(s). Reasons why the incidents were not reported were because the respondents were embarrassed by the situation (20\%), they didn't feel that someone would listen (37\%), they felt that other people have worse problems (25\%), they didn't know where to go for help (16\%) or due to a language barrier (2\%). The types of harassment or discrimination encountered are shown in Figure~\ref{fig:harassment}. Respondents who chose {\em other} cited discrimination based on nationality, discrimination based on hierarchy, discrimination due to having kids (or not), language discrimination  and discrimination due to research subject. Whilst the data is incomplete for the gender of respondents having experienced gender discrimination, 32 were female, 2 male and 0 other.

 We asked how often people felt overwhelmed by their situation at work over the course of the last year. 10\% felt overwhelmed all of the time and 34\% had often felt overwhelmed. Just 6\% had not felt overwhelmed at all over the last year. Of those that felt overwhelmed all of the time, none of them were permanent staff. 43\% of all respondents agreed that they were happy with their health and well-being.

\begin{figure}[ht!]
 \centering
 \includegraphics[width=0.7\textwidth,clip]{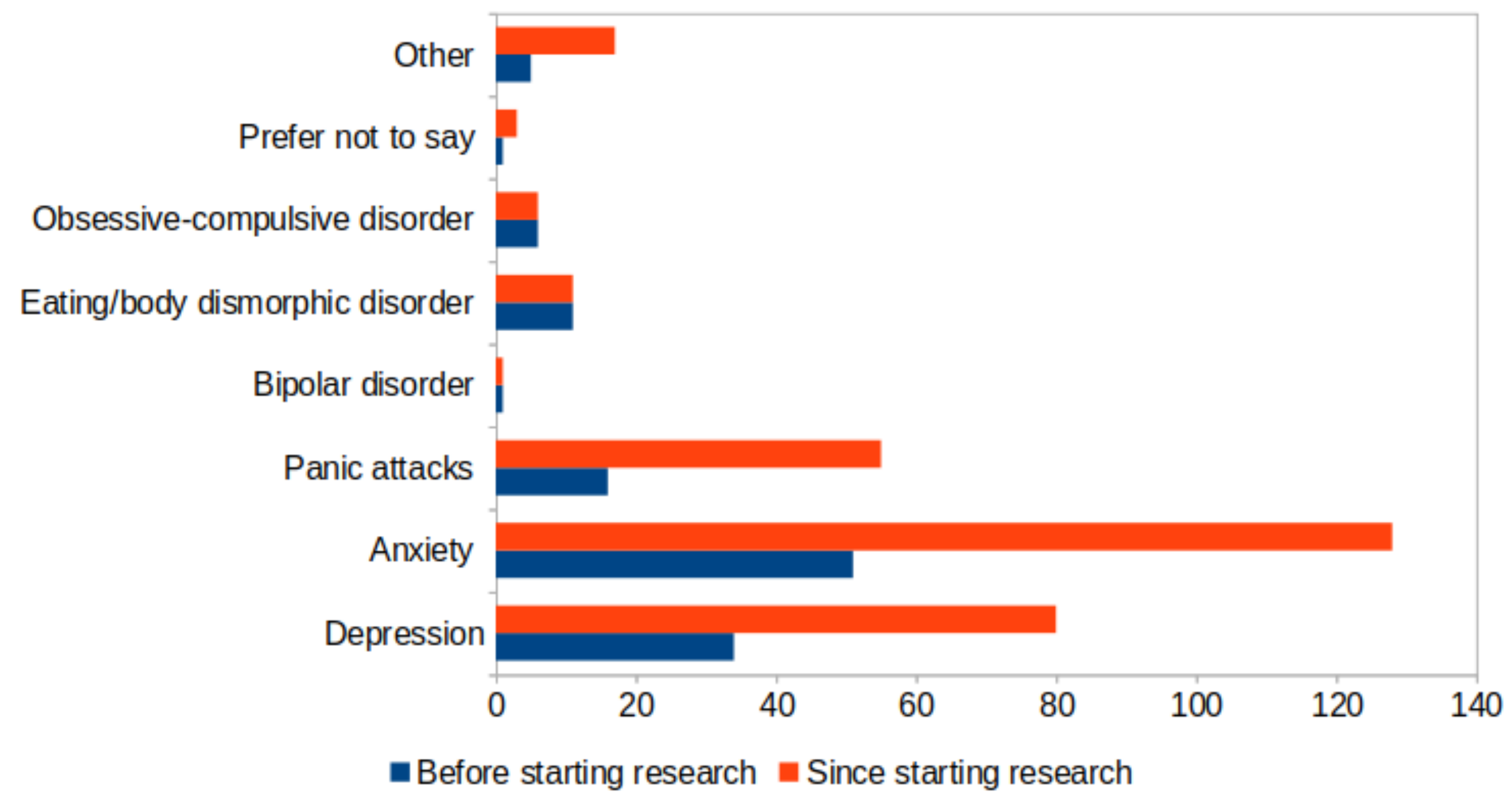}      
  \caption{The nature of the different mental health issues suffered by respondents and the number of people experiencing them, before starting in astrophysics research (blue) and since starting in the domain (red). Respondents who chose {\em other} cited post-traumatic stress syndrome, burnout, gender dysphoria, sleep disorder and imposter syndrome.
}
  \label{fig:MentalHealth}
 \end{figure}

 Concerning mental health, 20\% of respondents stated that they suffered from depression or other mental health problems before starting in research, however since starting their career, 44\% of respondents suffer from depression or other mental health problems. The types of health issues suffered by the respondents both before starting research in astrophysics and since starting are given in Figure~\ref{fig:MentalHealth} . 41\% have sought help for these issues. The other 59\% have not sought help because they were embarrassed by the situation (15\%), they didn't feel that someone would listen (13\%), they felt that other people have worse problems (51\%), they didn't know where to go for help (12\%) or due to a language barrier (not speaking the local language, 9\%). The problems experienced since starting astrophysics have caused some of the respondents to turn to alcohol abuse (23 people), drug abuse (8 people), other substance abuse (2 people), disordered eating (32 people), self-harm (9 people) and one suicide attempt in an aim to deal with their problems. 

  We also questioned about possible solutions. 81\% of respondents agreed that PhD and post-doc supervisors should get training in supervision and 73\% agreed that academic staff in general should get training in mental health issues, while 76\% felt that more discussion of well-being should take place.  Other suggestions from respondents included, longer temporary contracts ($>$ 3 years), mentoring or general help to construct a future either in or out of research, improving relations between permanent and temporary staff, improving communication by having more open discussion on working hours, racism, sexism, etc, providing a realistic outlook about research jobs etc before the PhD, improving the welcome and provision of information for new hires, providing information on how to access psychological support (in English), improving transparency in decisions made and minimising (or helping with) administrative tasks.

Finally, we also asked about the most positive experiences in astrophysics research. These were plentiful and included: publishing work, international connections, international recognition, launching satellites/instruments, making new discoveries, attending international conferences, working abroad, getting a PhD, doing outreach, teaching, collaborating, interdisciplinary work, meeting people, learning, supervising, receiving constructive feedback, having scientific discussions, observing, changing domains, having great ideas, having proposals accepted, promotions, press releases and sharing success and contributing to big projects.

\section{Discussion}

Whilst the majority of respondents were male, the percentage of female respondents (see Section~\ref{sec:results}) was far in excess of the number of female astrophysicists in France ($\sim$25\%). This is a well known phenomenon, where more women than men respond to online surveys e.g. \cite{cull05}. Women may also feel more concerned by the topics discussed, enhancing the percentage of respondents, as more women than men have suffered from gender discrimination (see Section~\ref{sec:results}) and more of the women stated that they suffered from depression and/or mental health issues (54\% of the women that responded) compared to 34\% of men that responded, similar to numbers found by \cite{evan18}. 100\% of those who replied to the gender question as 'other' also stated that they suffered from depression and/or mental health issues.

68\% of respondents worked significantly in excess of the legal number of hours per week prescribed in France, primarily due to external pressure. These were almost all non-permanent staff, the same group that felt that they were either often or constantly overwhelmed. The proportion of men and women working more than the number of legal hours was similar. It is clear that working in excess of 90 hours per week leaves no time for any other activity during the week outside of work, sleeping and eating, which is clearly unhealthy. However, only a third of those working long hours were the same people that felt that they did not get enough sleep, indicating that it was often not long working hours that reduced sleep. Stress could be a factor preventing people from sleeping. The pressure felt to publish and/or be competitive/obtain a permanent position were frequently cited in the reasons for working long hours. The figures indicate that temporary staff are working longer hours than permanent staff, putting them more at risk of poor (mental) health.

One of the major sources of stress is the precarity in astrophysics research, as seen from the number of people that plan to leave academia, but would stay, if they could be guaranteed a job. Precarity was also the most highly cited difficulty encountered in astrophysics, along with related issues such as geographical instability. However, a fifth of respondents would not definitely stay in academia even if they were guaranteed a permanent job. Of those that are definitely planning to leave, one third had experienced harassment or discrimination in astrophysics or had experienced mental health issues since starting, which could explain part of their motivation. A further third is due to precarity issues, leaving a final third who are leaving for other reasons.

The number of people suffering with mental health issues since starting their career in astrophysics has more than doubled compared to prior to starting their career. This is a startling rise, but numbers recorded are commensurate with the study of graduate students in biosciences at the University of Berkeley \citep{evan18} only slightly higher than in other studies e.g. \cite{wool19} who polled $>$6000 doctoral researchers from all subjects in various countries around the world or the study by \cite{auer18} carried out by the {\em World Health Organisation} that revealed that 31\% of students showed signs of major depression, general anxiety disorder or panic disorder etc, in the previous 12 months.  It is clear that there is a major problem, not just in France, but all over the world and this needs to be resolved to restore people's mental health and well-being. The problem needs to be tackled at the source and many of the suggestions proposed or agreed on by the respondents could go someway to help, namely training PhD and post-doc supervisors in supervision and training academic staff in mental health issues, along with more open discussion on well-being,  working hours, racism, sexism, sexual harrassment, bullying, etc. Lengthening temporary contracts and providing mentoring or general help to construct a future either in or out of research could also help (see Section~\ref{sec:recs}).

Finally, this survey was planned for release on March 30th 2020, but was delayed for a year because of the outbreak of the COVID-19 pandemic. It is clear that the unprecedented situation over the year before releasing the survey has added stress to everyone's lives and thus had an impact on well-being and mental health. It is difficult to fully decorrelate the impact of the COVID-19 pandemic on the long-term well-being. However, has evident from other publications \citep[e.g.][]{wool19,auer18} and our own observations, suffering in (early career) research has been prevalent for many years before the COVID-19 pandemic outbreak and the pandemic appears to have simply exacerbated existing problems.



\section{Recommendations}
\label{sec:recs}

Following the survey and numerous discussions amongst the authors of this paper, a dedicated workshop on the issue at the 2021 French National Astronomy meeting, along with further discussions with other groups which aim to improve well-being in astrophysics, we have drawn up a list of recommendations for institutes, masters programmes/doctoral schools and gouverning bodies. Some of the points may seem minor, but implementing the recommendations would go a long way to helping people feel included, accepted and happy in their job. We are aware that some of these recommendations have been put into place in some institutes, or some specific areas, but applying them across the board could help significantly to improve the day to day life of our colleagues.

During our discussions, it became clear that there is a lot of help and information already available, but it is often dispersed, making access complicated. We propose a single, highly visible webpage (e.g. as a part of the SF2A webpages) which regroups all of the relevant information (well-being, support for harassment, mental health support, etc) so colleagues and organisations know where to locate it. In addition, a working group is being put into place to help maintain the webpage and help implement current and future recommendations.

\subsection{Recommendations at the institute level}

A common source of stress for younger colleagues on short term contracts is that they are frequently required to move institutes, often in different countries where they don't necessarily speak the local language, and this requires a lot of administrative activities with many papers to fill out, often available in the local language only. This makes a short and simple job, long and complicated. Providing paperwork in English would facilitate this aspect. Further, a lot of new recruits find that no provision for an office space, a computer or other basics have been organised at their arrival. Putting these things into place in advance and introducing the new arrivals to key people immediately, can make new arrivals feel welcome and appreciated. Further, providing information booklets (with information on opening a bank account, finding an appartment, organisation within the institute, how to access the intranet, mailing lists, etc) along with introductory meetings (introducing key people, key structures, etc) will also help their integration. Up to date information on the local website (in English) and mailing lists (with specific mailing lists for students and post-docs) would also facilitate integration. More (non-)scientific events should be organised to facilitate colleagues meeting and creating a team spirit. This would also help bridge the perceived gap between permanent and temporary staff. Institute messages should ideally be written in the local language as well as English to be inclusive, as part of an effort to reinforce inclusivity.

Many (international) mentor programmes exist, but institutes should also provide a mentoring programme to be able to provide local information (recruitment possibilities in the area, specificities for applying for local jobs, applying for funding etc). Whilst finding mentors in an institute with a broad knowledge of working in astrophysics within France and abroad as well as working in industry, for example, will be difficult to find, having a pool of mentors identified (and their expertise) would mean that colleagues will know who to consult about which issue. Regular or occasional mentoring should be provided. Open discussion of common problems encountered (e.g. imposter syndrome, well-being, hours worked, racism, sexism, etc) are recommended. This could be through a weekly or bimonthly coffee or dedicated seminars, for example.

Not all students and post-docs will stay in academia. More information should be provided to younger colleagues on moving to other domains and to remove the perceived stigma that leaving academia is shameful. Institutes should aim to keep contact with previous researchers that have moved into industry. They could be invited once a year, perhaps during the PhD day, to provide information on finding jobs outside of academia.

\subsection{Recommendations for masters programmes and doctoral schools}

Several respondents to the survey felt that they did not have a clear picture about their chances of getting a permanent job in astrophysics before deciding to start a PhD and as a result felt that they wouldn't have chosen such a career path if they had had a proper understanding of the requirements. Masters programme should therefore provide information to all students spelling out the difficulties, to allow students to take an informed decision, however, labouring the point can be a source of stress to others. Doctoral schools should also try to form alumni associations, if one does not already exist at that particular University, so that current students can contact previous students to get feedback and tips on staying in research or moving into industry. More talks/information should also be provided about bad practice (supervisors requiring replies to emails late at night or at the weekend, long working hours, etc). In general, we recommend that the training during Masters and PhD programmes should take into account the fact that the majority of the students will not work in academia over the long term.

\subsection{Recommendations for gouverning bodies}

To avoid the stress of frequently changing jobs, and losing friends, support networks, etc, post-doc positions could be increased in length (ideally beyond three years). Where possible, post-doc salaries should be harmonised and provide for an evolution of the salary over the time the post-doc is employed. This could be done for example through salary recommendations when applying for national funding, if it is not yet implemented.

Everyone working in astrophysics should be trained in well-being, harassment, discrimination etc. This will enable them to be aware of problems and be familiar with the language associated with these issues. This training should be obligatory and done using online courses with tests, which need to be passed every few years. A proto-type demonstrator is currently being drawn up, to provide a basis for a nationally deployed platform. Every institute should have a clearly identified person that is properly trained to deal with harassment. This person should not be the direct superior of anyone within the institute to avoid a conflict of interest.

\section{Conclusions}

Through this short paper, we aimed to share results from our survey and provide some discussion and recommendations that could help improve the day to day lives of colleagues. By being aware of the problems and implementing solutions, we can all help to improve our colleagues lives and make our environment a better place to work in and above all preserve the (mental) health of all.

\begin{acknowledgements}
We thank the people that answered the survey, those who will take on board what has been said and those that will help improve research careers in the future. We are grateful to Beno\^it Mosser for insightful discussion and helpful comments on this paper. We acknowledge {\it Framaforms, Topcat} and {\it Excel} in the production of this work.  We remember our colleagues that departed too soon, your memory will help us strive towards a better tomorrow.
\end{acknowledgements}

\bibliographystyle{aa}  
\bibliography{Webb_S00} 

\end{document}